%% file: main.tex
\documentclass{article}

\usepackage{arxiv}

\usepackage[utf8]{inputenc} 

\usepackage{amsfonts}       
\usepackage{nicefrac}       
\usepackage{microtype}      
\usepackage{lipsum}

\usepackage{color}
\usepackage{listings}
\usepackage{booktabs}
\usepackage{array}
\usepackage{graphicx}
\usepackage{longtable}
\usepackage{subfigure}

\usepackage{cite}
\usepackage{url}
\usepackage{fancyhdr}

\usepackage{soul}

\usepackage{float}
\usepackage{listings}
\usepackage{mdwmath}
\usepackage{mdwtab}
\usepackage{multirow}
\usepackage{multicol}
\usepackage{rotating}
\usepackage{setspace}
\usepackage[utf8]{inputenc}
\usepackage{lineno}
\usepackage{listings}

\usepackage{mdwmath}
\usepackage{mdwtab}
\usepackage{multirow}
\usepackage{multicol}
\usepackage{array}
\usepackage{booktabs}

\usepackage{enumitem}
\usepackage{xspace}
\usepackage[export]{adjustbox}
\usepackage{graphicx}
\usepackage{color,soul}
\usepackage{rotating}
\usepackage{setspace}
\usepackage{amsmath} 
\usepackage{amssymb}
\usepackage{float}
\usepackage{xcolor}

\usepackage{hyperref}
\usepackage[numbers]{natbib}
\usepackage{url}

\title{A Taxonomy of Foundation Model based Systems through the Lens of Software Architecture}

\author{Qinghua Lu\thanks{Contact email: qinghua.lu@data61.csiro.au}, Liming Zhu, Xiwei Xu, Yue Liu, Zhenchang Xing, Jon Whittle\\
Data61, CSIRO, Australia
}

\begin{document}

\maketitle

\begin{abstract}
The recent release of large language model (LLM) based chatbots, such as ChatGPT, has attracted huge interest in foundation models. It is widely believed that foundation models will serve as the fundamental building blocks for future AI systems. 
As foundation models are in their early stages, the design of foundation model based systems has not yet been systematically explored. There is limited understanding about the impact of introducing foundation models in software architecture. Therefore, in this paper, we propose a taxonomy of foundation model based systems, which classifies and compares the characteristics of foundation models and design options of foundation model based systems. Our taxonomy comprises three categories: the pretraining and adaptation of foundation models, the architecture design of foundation model based systems, and responsible-AI-by-design. This taxonomy can serve as concrete guidance for making major architectural design decisions when designing foundation model based systems and highlights trade-offs arising from design decisions.

\end{abstract}

\textbf{Key terms - } Software architecture, foundation model, responsible AI, large language model, LLM, taxonomy, ChatGPT, GenAI

\section{Introduction}

The release of ChatGPT~\footnote{\label{ChatGPT}https://openai.com/blog/chatgpt}, Bard~\footnote{\label{Bard}https://bard.google.com}, and other large language model (LLM)-based chatbots has drawn huge attention on foundation models worldwide. Foundation models, such as LLMs, are pretrained on massive amounts of data and can be adapted to perform a wide variety of tasks and significantly improve productivity~\cite{bommasani2021opportunities}. With numerous projects already underway to explore their potential, it is widely predicted that foundation models will serve as the fundamental building blocks for many future AI systems. 

However, the design of foundation model based systems is still in an early stage and has not yet been systematically explored. There is limited understanding of the impact of introducing foundation models into software architecture. This includes the pretraining and adaptation of foundation models, the architectural role of foundation models in software systems, ways to communicate with foundation models, connecting foundation models  with internal data, etc. Many reusable solutions have been proposed to tackle various challenges in designing foundation model based systems, which motivates the creation of a design taxonomy for foundation model based systems.

On the other hand, the black box nature and the rapid advancements in foundation models have raised significant concerns about AI risks~\cite{van2023chatgpt}. 
There are unique challenges in implementing responsible AI for foundation model based systems, which requires responsible-AI-by-design solutions~\cite{lu2023responsible} for foundation model based systems. A key challenge is accountability, which becomes more complex with multiple stakeholders involved. This includes the foundation model based system owner, the foundation model provider, and various external tools and services providers. To ensure accountability, it is essential to design mechanisms for traceability. Another major challenge is trustworthiness of both the final results and intermediate process, which is important for aligning with human goals and fulfill trustworthiness criteria. 
Additionally, the potential misuse of foundation model based systems poses a considerable challenge. This requires continuous risk assessment to ensure the decisions and behaviour of foundation model based systems are trustworthy and responsible.

Taxonomies have been used in the software architecture community to gain a deeper understanding of existing technologies~\cite{mehta2000towards,gorton2015architecture}. By categorising existing work into a compact framework, taxonomies enable architects to explore the conceptual design space and make rigorous comparisons and evaluations of different design options. 
Therefore, in this paper, we present a taxonomy that defines categories for classifying the key characteristics of foundation models and design options of foundation model based systems. The taxonomy is structured into three categories: the pretraining and adaptation of foundation models, the architecture design for foundation model based systems, and responsible-AI-by-design. The development of this taxonomy is grounded in an extensive review of relevant literature, supplemented by insights obtained from our project experience. The taxonomy serves as guidance to help software architects make rational decisions in designing foundation model based systems.


The remainder of the paper is organised as follows. Section 2 discusses related work. Section 3 introduces the methodology of this study. Section 4 presents the design taxonomy. Section 5 analyses the threats to validity. Section 6 concludes the paper.

\section{Background and Related Work}
OpenAI launched ChatGPT~\footref{ChatGPT} in November 2022, a foundation model powered conversational AI product that has gained over 100 million users within two months of its release. This triggered an arms race among big tech companies to develop foundation model based generative AI (GenAI) product. Google responded with Bard~\footref{Bard}, its own conversational GenAI products. By February 2023, Microsoft had already integrated GPT-4 into its search engine, Bing. Another type of notable GenAI products is text-to-image generators, such as DALL-E~\footnote{https://openai.com/product/dall-e-2}, Stable Diffusion~\footnote{https://stablediffusionweb.com}, and Midjourney~\footnote{https://www.midjourney.com}. Additionally, there are domain-specific foundation models which can be applied in fields such as finance~\cite{wu2023bloomberggpt}, medicine~\cite{moor2023foundation}, and climate~\cite{nguyen2023climax}.

There is growing interest in integrating foundation models into existing software systems or building new systems. To meet specific requirements and ensure model accuracy, many studies explore techniques for adapting foundation models for downstream tasks, such as using domain-specific data~\cite{jablonka2022gpt} or in-context learning~\cite{xie2021explanation}.  
There are also many efforts on prompt engineering, including various prompt patterns~\cite{schmidtcataloging} and prompt engineering tools used for developing foundation model based systems. Langchain~\footnote{https://www.langchain.com} is the most recognised prompt engineering tool. However, there are some practical challenges with Langchain, including complex software architecture and difficulties in debugging and maintenance. This has led to the emergence of competitors, such as AutoChain\footnote{https://github.com/Forethought-Technologies/AutoChain}, MiniChain\footnote{https://github.com/srush/MiniChain}, and Microsoft Prompt Flow\footnote{https://github.com/microsoft/promptflow}. These tools primarily focus on assembling task instructions for foundation models. 
However, there remains a lack of systematic research on software architecture design for foundation model based systems, such as major design decisions in the design process and their impact on quality attributes.

\begin{figure} [tbp]
\centering
\includegraphics[width=0.5\columnwidth]{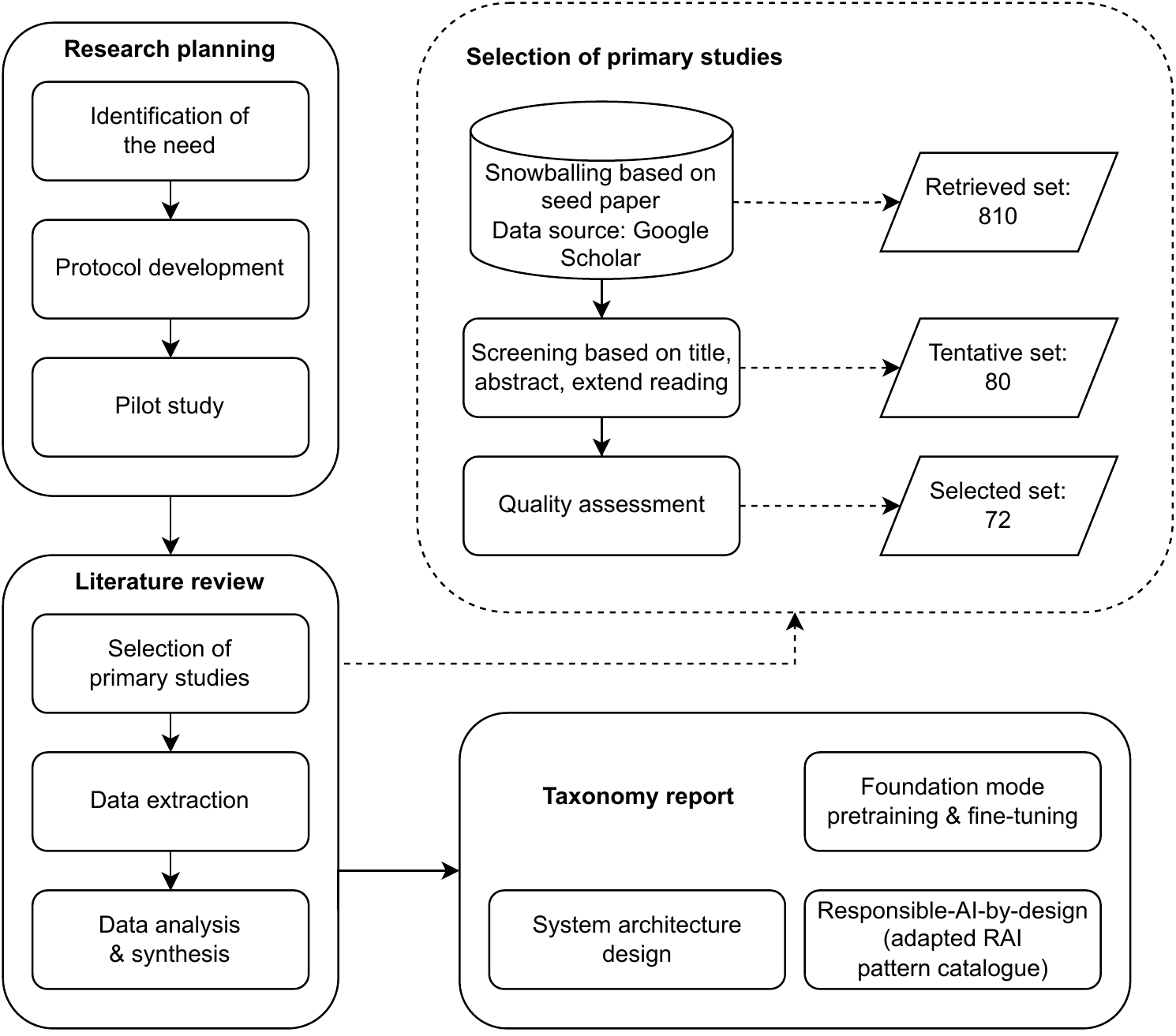}
\caption{Methodology.} \label{fig:protocol}
\vspace{-2ex}
\end{figure}

Although foundation models have huge potential to improve productivity, they also pose significant AI risks, such as hallucinations,  lack of source evidence, potential for user manipulation, dissemination of fake news. In April 2023, a petition~\footnote{https://futureoflife.org/open-letter/pause-giant-ai-experiments/} signed by over 1000 well-known AI researchers and practitioners from around the world, including Elon Musk and Stuart Russell, called for a six-month pause in the development of foundation models larger than GPT-4, as well as the creation of a common set of mechanisms for responsible AI. 
Various mechanisms, such as regulation, standards, tools, are typically used to address responsible AI. Many of these mechanisms have been summarised as patterns, including governance patterns, process patterns, and product patterns~\cite{lu2022responsible}. However, there are limited studies specifically designed for the development of responsible foundation model based systems. 

\section{Methodology}
In this section, we introduce the methodology for developing this taxonomy. Fig.~\ref{fig:protocol} illustrates the overall research steps of this study. Despite the tremendous attention and interest that foundation model based systems are attracting  worldwide, there is a lack of guidance for designing and classifying these emerging systems. We formulated three research questions as follows, and adopted the methodology of systematic literature review\cite{GAROUSI2019101, keele2007guidelines} to investigate the state-of-the-art studies to answer these questions.

\begin{itemize}
    \item \textit{What are the pretraining and adaptation mechanisms for foundation models?} This question aims to understand the design options available for pretraining and tuning foundation models and their implications.

    \item \textit{What are the key architectural decisions in designing foundation model based systems?} This question is intended to identify the various architectural choices in designing foundation model based systems.

    \item \textit{How can responsible AI be implemented in the design of foundation model based systems?} This question focuses on the available design options for integrating responsible AI into the architecture of foundation model based system.
\end{itemize}


\begin{figure*} [tbp]
\centering
\includegraphics[width=0.9\textwidth]{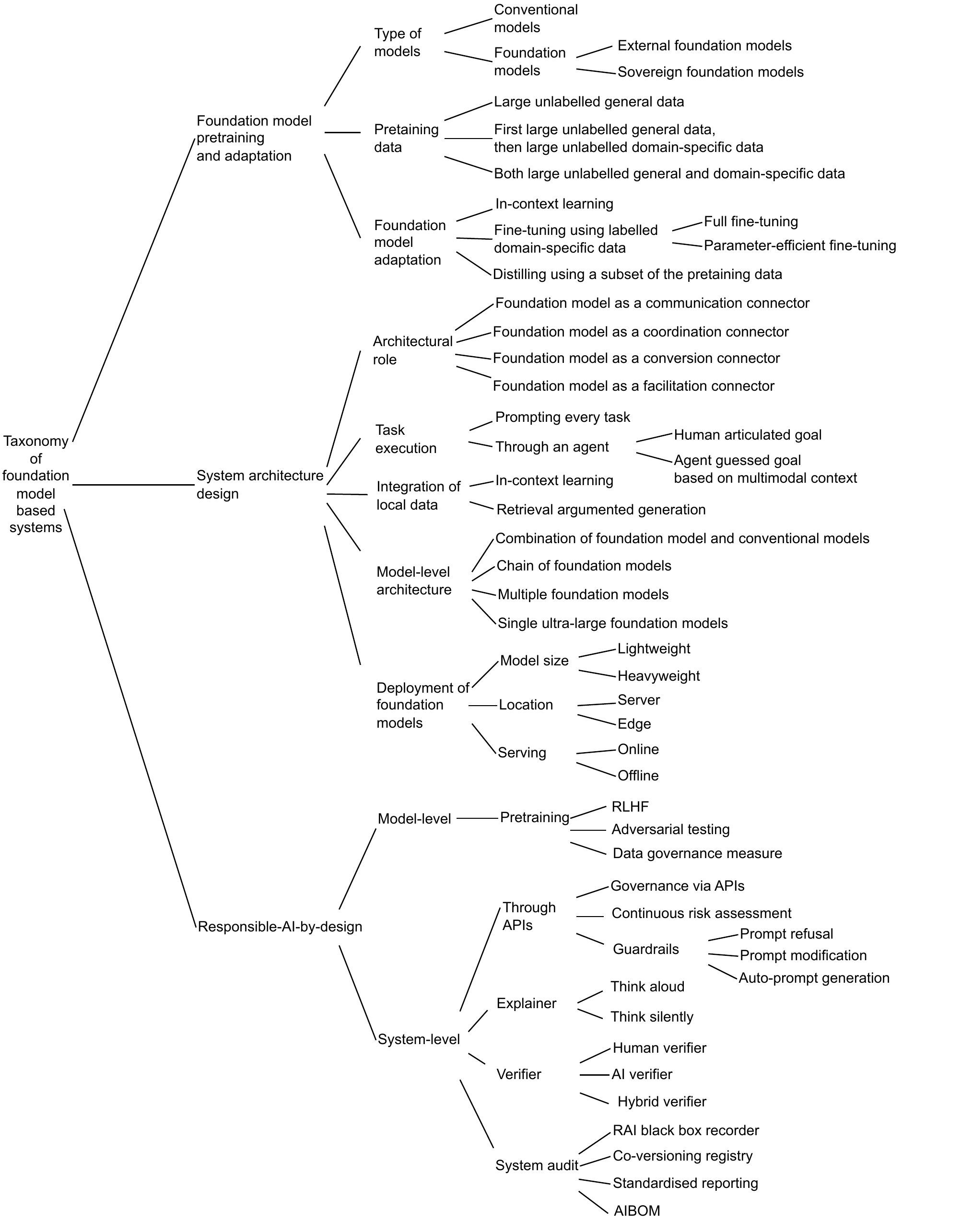}
\caption{Taxonomy of foundation model based systems.} \label{fig:taxonomy}
\vspace{-2ex}
\end{figure*}

\begin{figure*}
\centering
\includegraphics[width=0.55\textwidth]{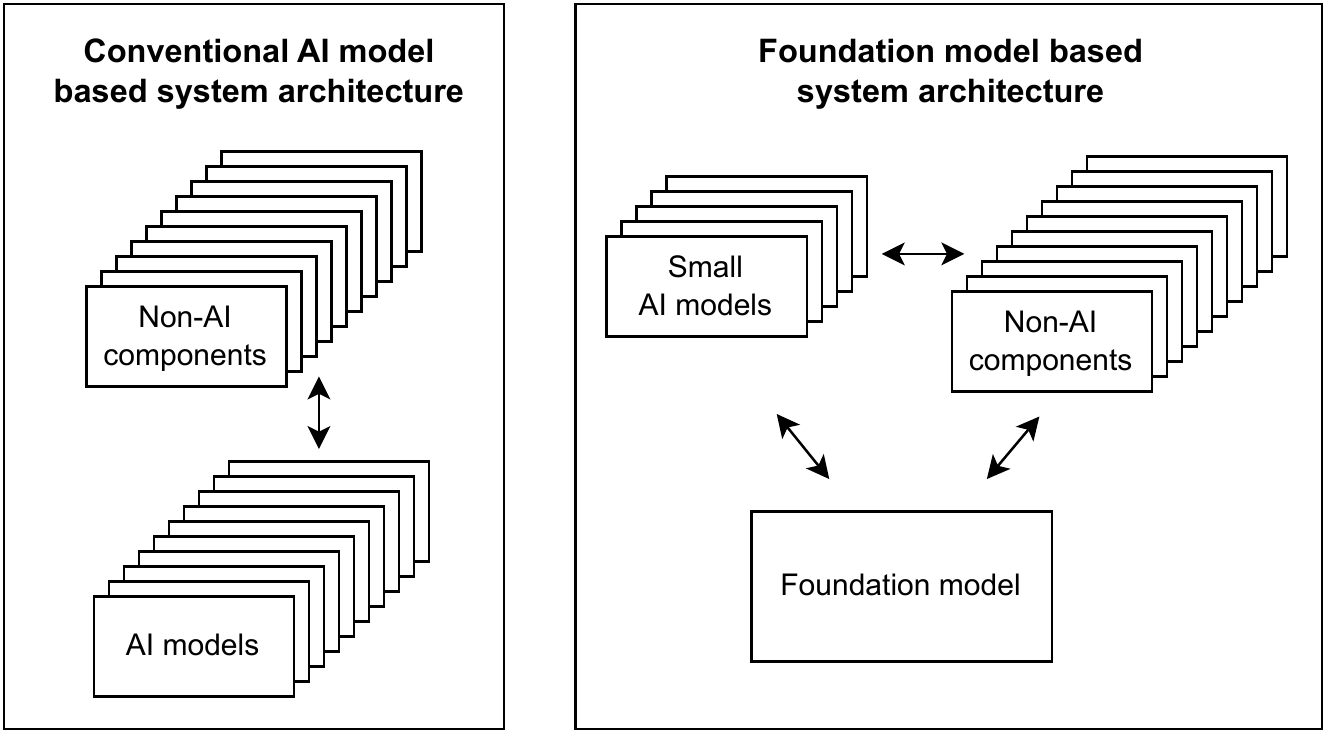}
\caption{Conventional AI models or foundation models.} \label{fig:cm-fm}
\vspace{-2ex}
\end{figure*}

To collect the primary studies for our literature review, we selected the seminal paper ``\textit{On the Opportunities and Risks of Foundation Models}''~\cite{bommasani2021opportunities} as our seed paper. This paper is notable for introducing the term ``foundation model'' and has been widely cited within the field of foundation models and foundation models. We initially retrieved 810 studies that cite the seed paper. After applying specific inclusion and exclusion criteria, 80 papers were identified for the study. The selection criteria are detailed as follows.

\textbf{Inclusion criteria:}

\begin{itemize}
    \item  A paper that discusses foundation model based systems.
    \item  A paper that discusses responsible AI  of foundation model.
\end{itemize}

\textbf{Exclusion criteria:}

\begin{itemize}
    \item Papers not written in English.
    \item Conference version of a study that has an extended journal version.
    \item Survey and review papers. 
    \item PhD/Master’s dissertations, tutorials, editorials, books.
\end{itemize}

Afterwards, we performed a quality assessment on the tentative set of literature. 72 primary studies were finally included for data extraction and synthesis. One researcher was responsible for extracting the answers, while three researchers engaged in the consultation and analysis process to establish the taxonomy based on the results. Specifically, we adapted some patterns from the responsible AI pattern catalogue~\cite{lu2022responsible} when building the Responsible-AI-by-design category. 

\section{Design Taxonomy}
In this section, we present a taxonomy that defines categories for building foundation models and integrating foundation models into the design of software systems. Our taxonomy is designed to help software architects and developers in comparing and evaluating foundation models, as well as enable research into design decision-making for building foundation models and foundation model based systems.

As illustrated in Fig.~\ref{fig:taxonomy}, our taxonomy is structured in three categories: the pretraining and adaptation of foundation models, architectural design of foundation model based systems, and responsible-AI-by-design. We discuss the characteristics of each design options and their impact on cost, accuracy, and responsible AI (RAI) related properties (including privacy, fairness, safety, security, explainability, transparency, contestability, accountability, human-centred values, and human, societal, and environmental wellbeings~\footnote{https://www.industry.gov.au/publications/australias-artificial-intelligence-ethics-framework/australias-ai-ethics-principles}).


\subsection{Foundation model pretraining and adaptation}
\subsubsection{Using conventional AI models or foundation models}\hfill\\
Whether to use conventional AI models or foundation models is a major architecture design decision when designing AI systems, as illustrated in Fig.~\ref{fig:cm-fm}. 
\textbf{Conventional AI models} are trained from scratch on a specific task using a dataset that is collected and labelled specifically for that task. 
Most current AI systems comprises conventional AI models and non-AI components, and have not integrated foundation models into the design. 
Building conventional AI models for various tasks can be costly, requiring extensive data gathering and labelling. Limited amounts of training data and computation resources available within organisations can lead to low accuracy of the models. 






On the other hand, \textbf{foundation models}, such as LLMs, are large AI models that are pretrained on massive amounts of broad data for general-purpose tasks such as language processing or image recognition, which can be then adapted to perform a wide variety of tasks through fine-tuning~\cite{bommasani2021opportunities}. Using a foundation model can reduce human labour for developing the AI components, 
as organisations either only need to pay for external foundation model usage fees based on the number of API requests made or share a single foundation model within the organisation. Fine-tuning can improve the accuracy of foundation models for downstream tasks. In addition, a foundation model can interact with other small AI models for specific tasks. However, managing the foundation model pipeline can be more complex for organisations, whether the model is pretrained externally or internally. There can be more RAI-related issues with foundation model outputs, especially given the large-scale deployment of foundation models. For example, the huge amounts of data used to train these models may be biased or include sensitive information. The external organisations may use the data collected from these models for their own purposes, such as improving their products or developing new features, which can raise privacy concerns.






\begin{figure*}[tbp]
\centering
\includegraphics[width=0.65\textwidth]{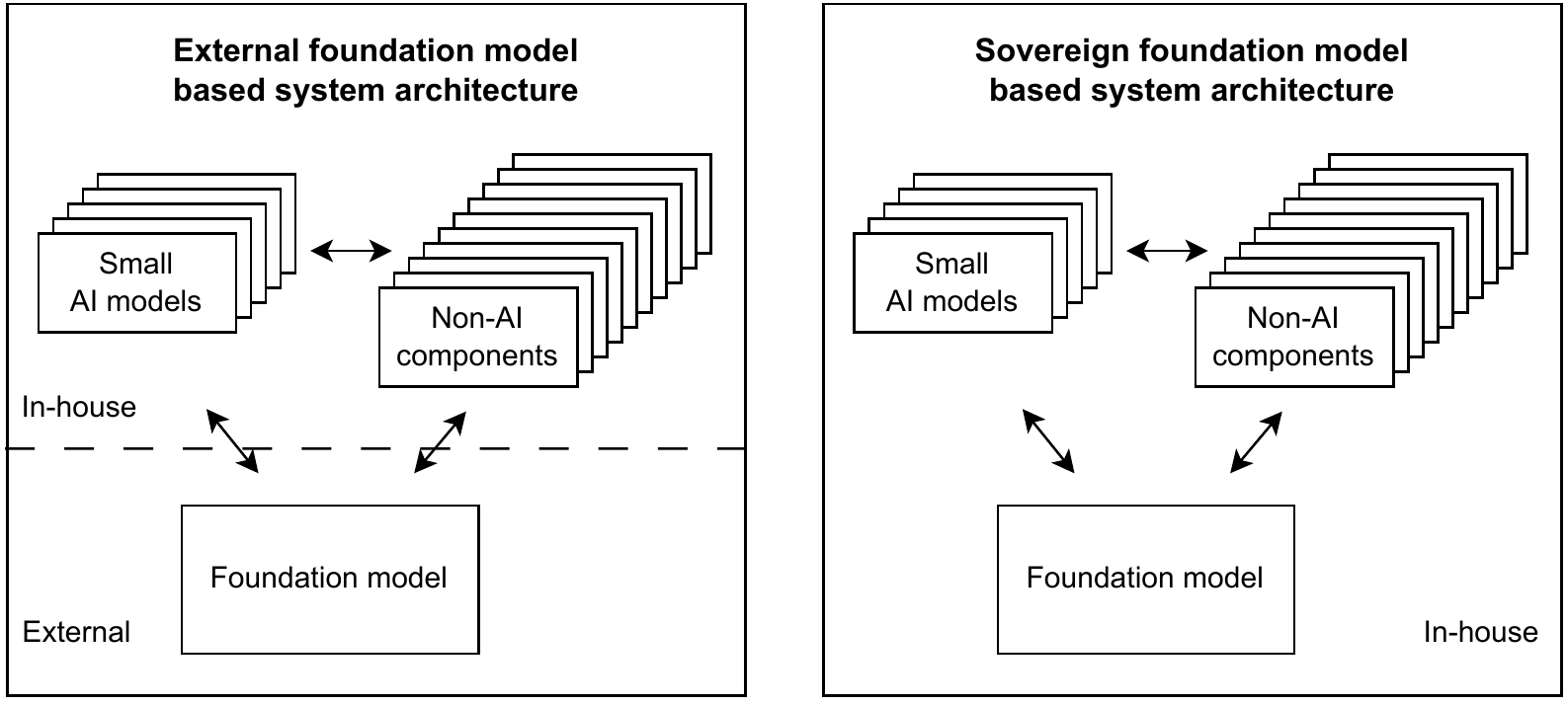}
\caption{External foundation models or sovereign foundation models.} \label{fig:ex-in}
\vspace{-2ex}
\end{figure*}

\subsubsection{Using external foundation models or sovereign foundation models}\hfill\\

As illustrated in Fig.~\ref{fig:ex-in}, one of the most critical decisions when designing the architecture of foundation model based AI systems is choosing whether to use a foundation model pretrained by an external organisation or build a sovereign foundation model in-house from scratch. 
Using an \textbf{external foundation model} can save human resources for model training, deployment and maintenance, 
as organisations only pay for API calls. Additionally, using a well-established foundation model can potentially result in higher accuracy and generalisability to a wider range of tasks. However, there can be more RAI-related issues as the organisations can only rely on limited means to ensure RAI qualities due to the black box nature of external foundation models. For example, privacy concerns can arise as organisations are uncertain whether their customers' data will be reused without acknowledgement.

On the other hand, some organisations may possess unique internal data and training a \textbf{sovereign foundation model} in-house from scratch can become their unique competitive advantage. Also, the sovereign foundation  model provides organisations with complete control over the model pipeline. The organisations can train the sovereign foundation model to meet specific needs and ensure RAI-related properties. The trained foundation model can be shared across different departments. However, this requires significant investments in terms of cost and resources, including data, computational power, and human resources. Moreover, it may take considerable expertise and time to train a foundation model that can achieve the desired accuracy and RAI-related qualities. For example, the UK recently announced initial start-up funding of £100 million to build its own sovereign foundation models~\footnote{https://www.gov.uk/government/publications/integrated-review-refresh-2023-responding-to-a-more-contested-and-volatile-world}.







\subsubsection{Pretraining data}\hfill\\

\begin{figure*} [tbp]
\centering
\includegraphics[width=0.7\textwidth]{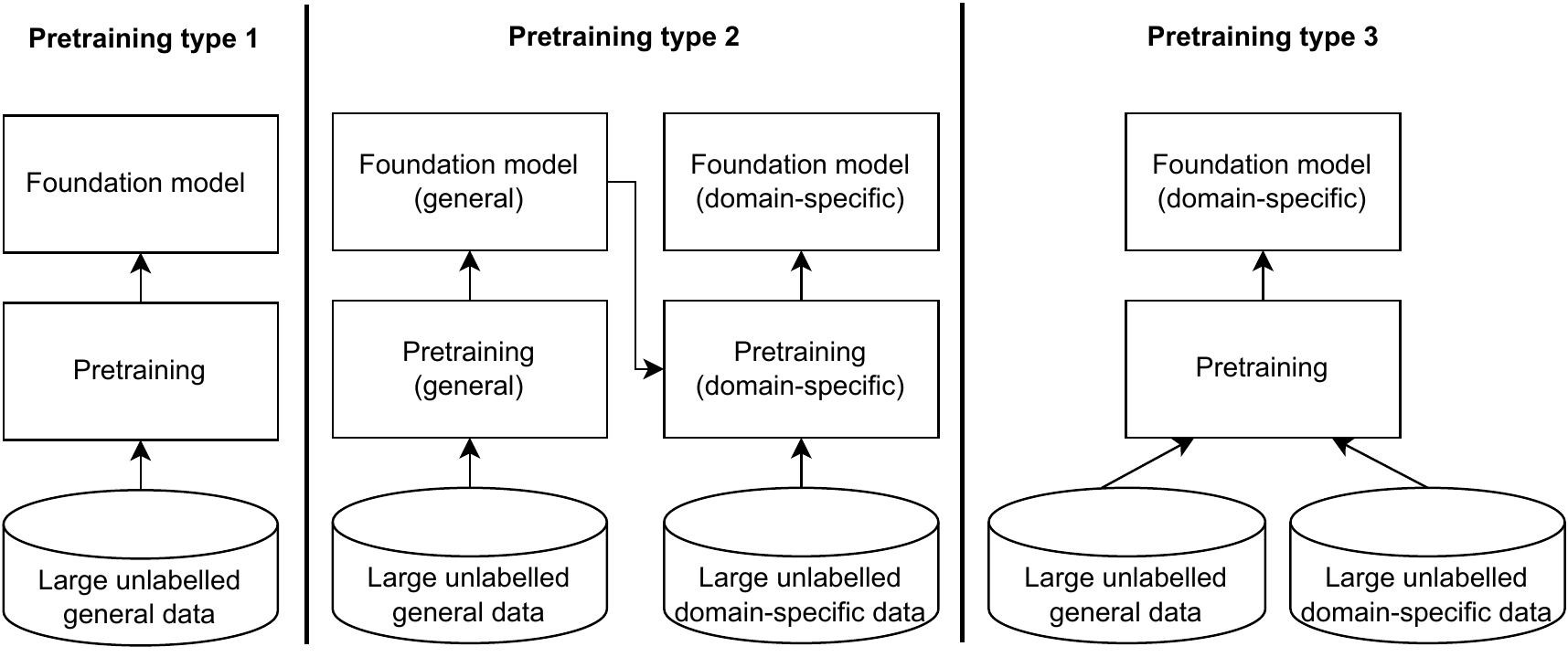}
\caption{Pretraining data.} \label{fig:pretraining}
\vspace{-2ex}
\end{figure*}

Fig.~\ref{fig:pretraining} illustrates three types of foundation models based on pretraining data. 
A foundation model (pretraining type 1) is typically pretrained by an organisation (e.g., big tech companies) using \textbf{large, unlabelled, and general data}, e.g., general text corpus. This foundation model can then be further pretrained by the same organisation or a different organisation using \textbf{large, unlabelled, domain-specific data} (pretraining type 2)~\cite{PPR:PPR661659}, e.g., public real-estate data. A foundation model (pretraining type 3) can also be pretrained by an organisation using \textbf{both large, unlabelled, general data and large, unlabelled, domain-specific data simultaneously}~\cite{lin2023medical}. While pretraining type 2 \& 3 require more cost and resources, they are capable of achieving better accuracy when performing domain-specific tasks and can improve RAI-related qualities (such as reliability). Pretraining type 3 may require less time and resources compared to pretraining type 2, since the pretraining process is carried out in a single step rather than sequentially. However, pretraining type 3 may require more complex infrastructure to manage different types of data, and may also require additional safeguards to ensure RAI.

\begin{figure*} [tbp]
\centering
\includegraphics[width=0.7\textwidth]{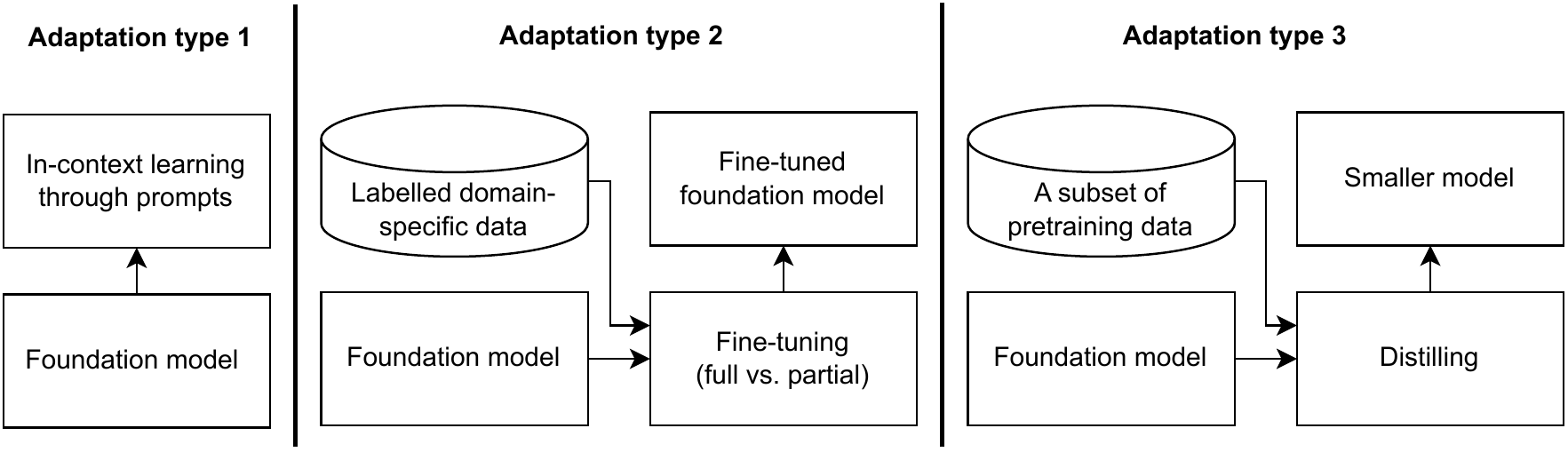}
\caption{Model adaptation.} \label{fig:adaptation}
\vspace{-2ex}
\end{figure*}







\subsubsection{Foundation model adaptation}\hfill\\

As shown in Fig.~\ref{fig:adaptation}, there can be different ways to adapt foundation models for downstream tasks. 
\textbf{In-context learning} (adaptation type 1) is to improve the accuracy of the foundation model by providing examples to perform a specific task. This allows the foundation model to perform a task more accurately without the need to tune any parameters. There are still accuracy and RAI issues with the model outputs. This is due to limited input-output examples size that can be used through prompts, which can result in the model generating outputs that lack the necessary context and understanding required for accurate inference.

\textbf{Fine-tuning the parameters of the foundation model using labelled domain-specific data} (adaptation type 2) can be done in two ways. Full fine-tuning retrains all the parameters, which is less feasible and extremely expensive (such as GPT-3 with 175B parameters). Another way is to employ parameter-efficient fine-tuning techniques, e.g., LoRA~\cite{hu2021lora} which reduces the number of training parameters by 10,000 times and decreasing GPU usage by threefold. 
If there is a need for improving the model's architecture and using pretraining  model weights directly is not feasible, \textbf{distillation} (adaptation type 3) becomes necessary. Distilling the foundation model is about training a smaller and more lightweight model to mimic the behavior of the larger and more complex foundation model using knowledge transfer techniques such as knowledge distillation, attention distillation, or parameter sharing. The distilled model is trained on a subset of the same data used to train the foundation model, but with a different loss function that encourages it to reproduce the output of foundation model. This can result in a more efficient and lightweight model that still maintains an acceptable level of performance.
However, fine-tuning or distilling the foundation model highly couple with the open source model (e.g., LLaMA~\cite{touvron2023llama}), which are not portable.



\subsection{Architecture design of foundation model based systems}
\subsubsection{Architectural role: foundation-model-as-a-connector} \hfill\\
In software architecture, software connectors are the fundamental building blocks of interactions between software components~\cite{mehta2000towards}, which can provide four services: communication, coordination, conversion, and facilitation.

\begin{figure} [tbp]
\centering
\includegraphics[width=0.49\columnwidth]{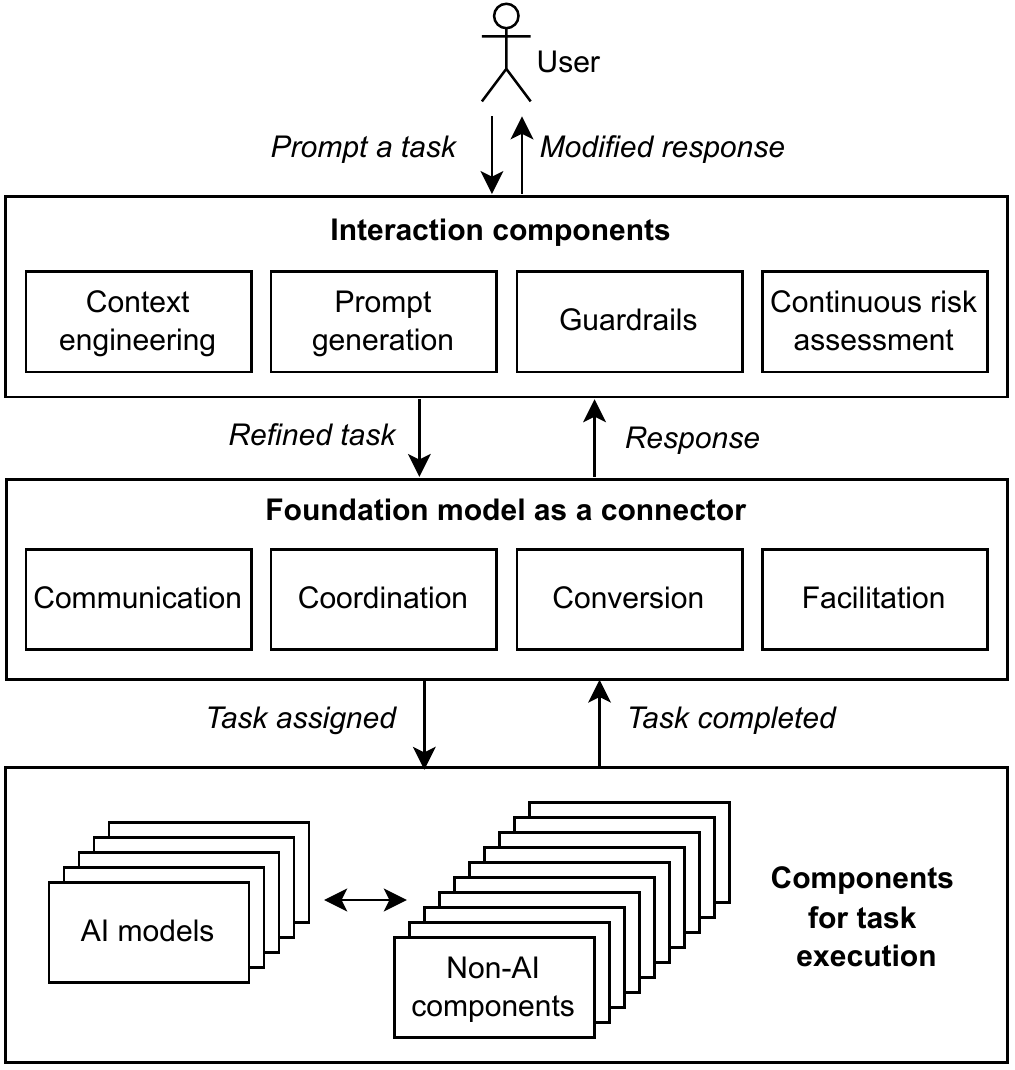}
\caption{Foundation-model-as-a-connector.} \label{fig:connector}
\vspace{-2ex}
\end{figure}

As shown in Fig.~\ref{fig:connector}, from an architectural perspective, a foundation model can be considered as a software connector. Once the task is prompted by a user, it can be further refined by interaction components for accuracy and RAI-related properties (see Section 3.3). The foundation model can play an architectural role as the following connector services to connect with other non-AI/AI components provided by the organisation: 

\begin{itemize}
    \item \textbf{Communication connector}: Foundation models such as LLMs can serve as a communication connector that enables the transfers of data between software components. For example, an LLM can be employed to receive task text descriptions from users and extract meaning and intention from the text. The extracted task information can then be transferred to other components for further processing, such as sending to an AI model or a robotics system to perform a specific task~\cite{shen2023hugginggpt, driess2023palm}.

    \item \textbf{Coordination connector}: Different software components can coordinate their computations through a foundation model. For example, an LLM can be used to plan a complex task or a workflow, which coordinates the planning, selection, and cooperation of multiple AI models through a text interface~\cite{shen2023hugginggpt, hong2023metagpt}. The LLM first needs to decompose the task into a set of subtasks and decide dependencies and execution order for these subtasks. Then the LLM needs to obtain the model description (e.g., functionality, architecture, domains, etc) and match the tasks and models.

    
    \item \textbf{Conversion connector}: Foundation models can function as an interface adapter for software components that use different data formats to communicate with each other. For example, an LLM can analyze text task descriptions from users and parse it into machine-readable template (e.g., task ID, type, dependencies, arguments, etc.) for executing by an AI model~\cite{shen2023hugginggpt}. 
    \item \textbf{Facilitation connector}: Foundation models can be integrated as a facilitation connector to optimise interactions between components. For example, an LLM can be employed to maintain chat logs, determine whether to run some commonly-used and time-consuming models locally, manage resource dependencies between tasks in task execution stage, and summarise the task execution process and inference results ~\cite{shen2023hugginggpt}. 
    
\end{itemize}

\begin{figure*} [tbp]
\centering
\includegraphics[width=0.75\textwidth]{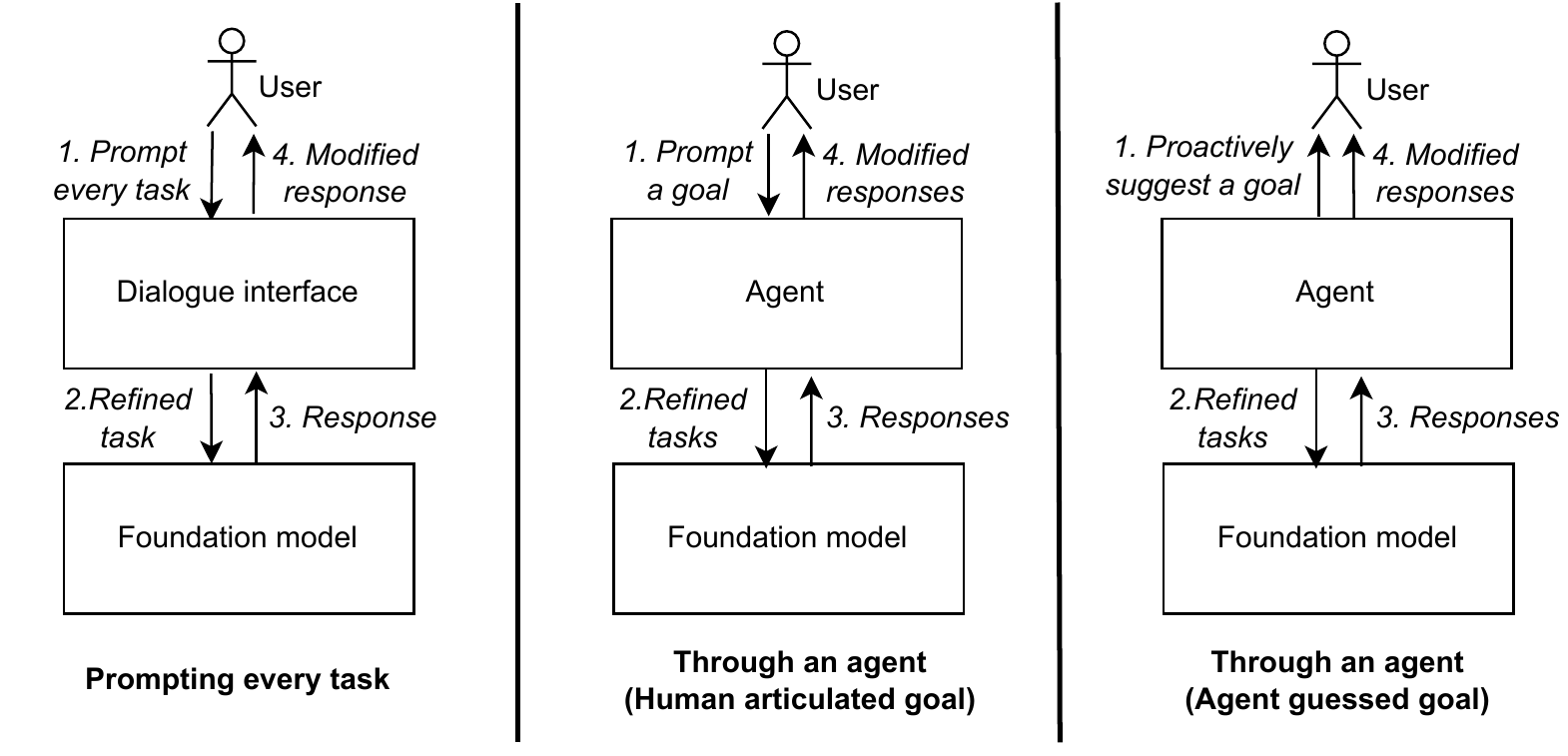}
\caption{Task execution with foundation models.} \label{fig:agent}
\vspace{-2ex}
\end{figure*}

\subsubsection{Task execution with foundation models} \hfill\\
The dialogue interface receives user-provided prompts for the foundation model to respond to, such as instructions and questions. Prompt engineering is a process of creating prompts that guides the output of a foundation model to meet specific needs. Fig.~\ref{fig:agent} illustrates three design options for task execution with foundation models. Using the default dialogue interface can be inefficient for complex tasks or workflows, as users may need to prompt every single step. To ensure accuracy and RAI-related properties, prompt patterns are often applied to guide the output of foundation models. There are various prompt patterns available, such as zero/one/few-shot prompt, retrieval/internet-augmented prompt, chain of thought, think aloud, bot team, negative prompt, multiple choice prompt, etc. When selecting the prompt patterns, users need to consider a variety of factors, such as the system's goal, target users, context, and specific tasks being performed. Each prompt pattern may require different costs and have varying levels of complexity. 







An \textbf{autonomous agent}, such as Auto-GPT~\footnote{https://github.com/Significant-Gravitas/Auto-GPT}, BabyAGI~\footnote{https://github.com/yoheinakajima/babyagi}, AgentGPT~\footnote{https://github.com/reworkd/AgentGPT}, can be used to reduce the need for prompt engineering and the associated cost. With an autonomous agent, users only need to prompt the overall goal, and the agent can break down the given goal into a set of tasks and use the other software components, the internet and other tools to achieve those tasks in an automatic way. Alternatively, 
the agent can go beyond the explicit user text prompt and anticipate the user's goals by understanding the user interface (UI) of relevant tools and human interaction. This is facilitated through the analysis of multimodal context information, including screen recording, mouse clicks, typing, eye tracking, document annotations and notes.  

However, replying solely on an autonomous agent can lead to issues with accuracy, as the agent may not fully understand the users' intentions. On the other hand, hybrid agents, such as GodMode~\footnote{https://godmode.space} and AI chain~\footnote{https://aichain.online}, involve users in the loop to confirm the plan and provide feedback. 
A continuous risk assessor can continuously monitor and assess AI risk metrics to prevent the misuse of the AI systems and to ensure the trustworthiness of the systems.


\begin{figure*} [tbp]
\centering
\includegraphics[width=0.8\textwidth]{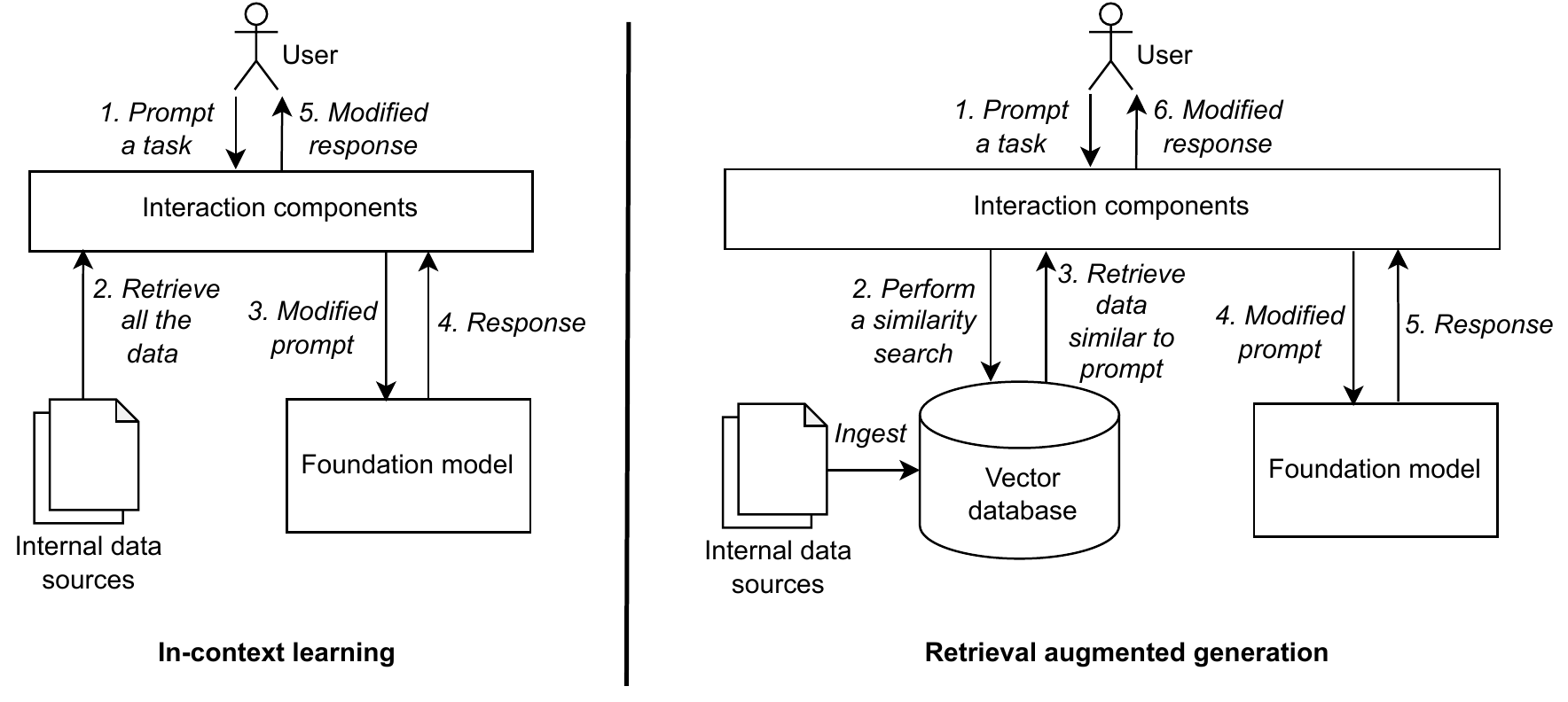}
\caption{Integration of local data} \label{fig:vector}
\vspace{-2ex}
\end{figure*}

\subsubsection{Integration of internal data}\hfill\\
It is often challenging to leverage organisation's internal data (e.g., documents, images, videos, audios, etc.) to improve the accuracy of foundation models, as foundation models are often inaccessible or expensive to retrain or fine-tune. To address this, there are two design options to consider, as demonstrated in Fig.~\ref{fig:vector}. The first design option is \textbf{in-context learning}, which integrates the internal data through prompts. However, there is a token limit for the context, e.g., the latest version of GPT-4 has a maximum limit of 128k tokens, equivalent to approximately 300 pages of text~\footnote{https://openai.com/blog/new-models-and-developer-products-announced-at-devday}. It is difficult to do a prompt with context data larger than the token limit. 
Alternatively, another design option is \textbf{retrieval augmented generation (RAG)}, which uses a vector database (such as Pinecone~\footnote{https://www.pinecone.io} and Milvus~\footnote{https://milvus.io}) for storing the personal or organisational internal data as vector embeddings. These embeddings can be used to perform similarity searches and enable the retrieval of internal data that are related to specific prompts. Compared to the direct integration of internal data through prompting, using a vector database is more cost-efficient and can achieve better accuracy and RAI-related properties.


\subsubsection{Model-level architecture}
\begin{figure*} [tbp]
\centering
\includegraphics[width=0.7\textwidth]{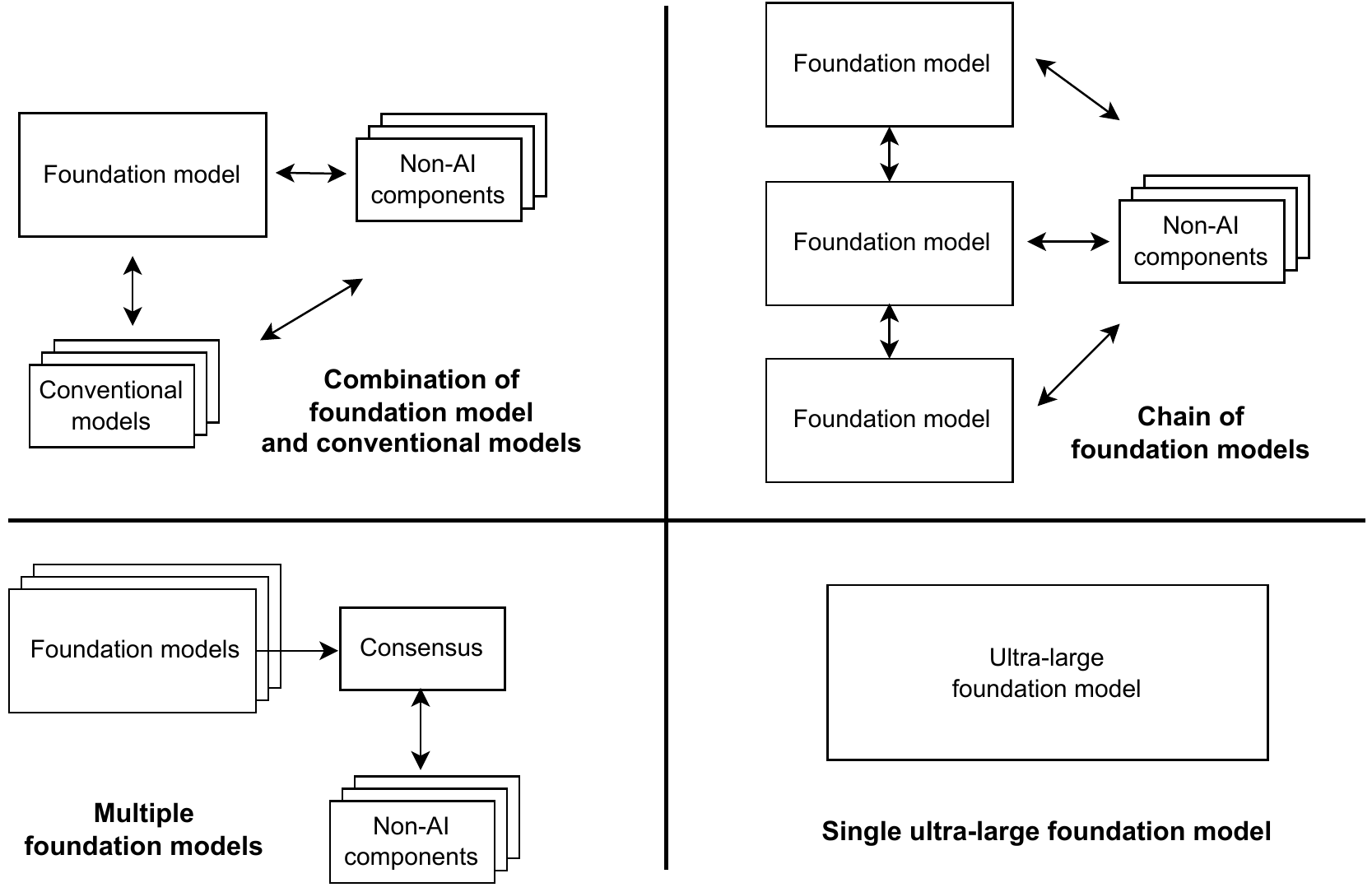}
\caption{Model-level architecture.} \label{fig:model}
\vspace{-2ex}
\end{figure*}

\hfill\\
Fig.~\ref{fig:model} shows four design options for the architecture of future foundation model based systems. The first design option is a \textbf{combination of the foundation model and conventional models}. 
The second design option is \textbf{a chain of foundation models}, which is modularised architecture, such as Socratic Models~\cite{zeng2022socratic}. This architecture relies on a few foundation models that are chained together and a limited number of AI and non-AI components to perform tasks (e.g., through language-based interactions) without requiring additional training or fine-tuning. The inference for a task-specific output is jointly performed by multimodal interactions between the independent foundation models, such as LLMs, visual LLMs and audio LLMs. Those foundation models can be connected via APIs with external AI or non-AI components that offer additional capabilities or access to databases, such as robotic systems or web search engines. By using task formulation and multimodal interaction between independent models, the architecture can effectively leverage the capabilities of different foundation models and external AI and non-AI components. 
The third design option is to \textbf{employ multiple foundation models in parallel} to perform the same task or enable a single decision.
The fourth design option is a \textbf{monolithic architecture}, which only contains a single big foundation model capable of performing a variety of tasks by incorporating different types of sensor data for cross-training. An example of this type of architecture is PaLM-E~\cite{driess2023palm}, which is used for performing language, visual-language, and reasoning tasks. With the rapidly growing capabilities, in this type of architecture, no external software components might be required.

\subsubsection{Deployment of foundation models}

When deploying foundation models, one important factor is the size of the foundation model, which can range from lightweight to heavyweight. A lightweight foundation model typically has a smaller number of parameters, making it suitable for edge devices. On the other hand, heavyweight models have a larger number of parameters, providing more powerful capabilities but requiring more computational resources.
In terms of the \textbf{location}, foundation models can be deployed on a server or at the edge. When a model is deployed on a server, there is no maintenance cost for the users. Deploying a foundation model at the edge makes the processing closer to the data source, reducing latency and allowing for real-time inference.
In terms of the \textbf{serving mode}, foundation models can be deployed online or offline. 
Online deployment involves making API calls to a remote server for inference. This approach requires an active internet connection but allows for model governance. Offline deployment enables the model to be served directly on a device without relying on internet connectivity, making it suitable for scenarios where internet access is limited or when real-time or offline inference is required.
For example, PaLM 2~\footnote{https://ai.google/discover/palm2} is available in a range of sizes, Gecko, Otter, Bison and Unicorn. Geckos is a lightweight PaLM 2 that can be deployed on mobile devices and serve offline.

\subsection{Responsible-AI-by-Design}

\begin{figure*} [tbp]
\centering
\includegraphics[width=0.6\textwidth]{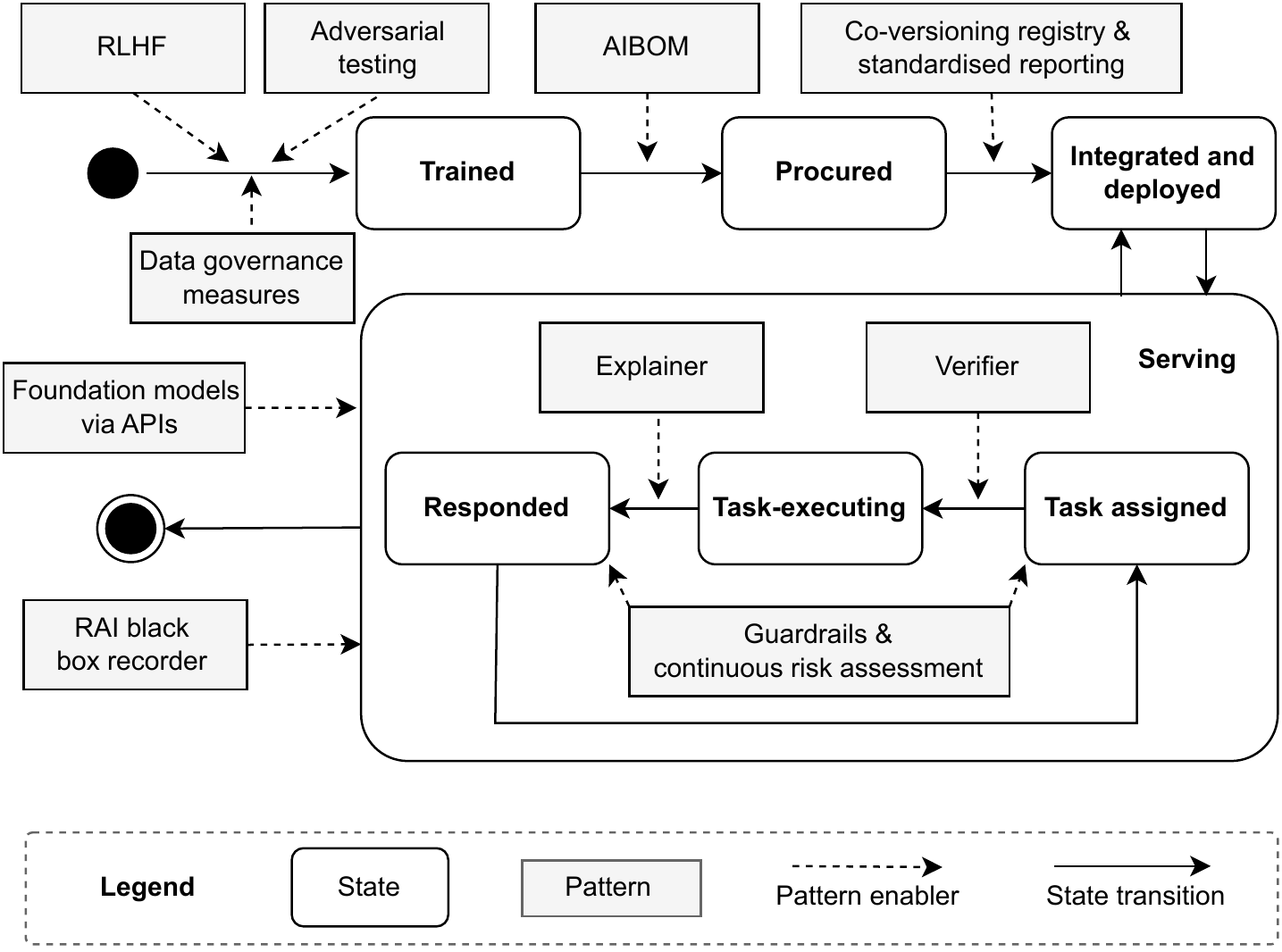}
\caption{Responsible-AI-by-Design.} \label{fig:rai}
\vspace{-2ex}
\end{figure*}

One key challenge of designing foundation model-based system is how to control the risks introduced by foundation models. Fig.~\ref{fig:rai} lists the model-level and system-level design options for responsible-AI-by-design. 

\subsubsection{Model-level risk control}\hfill\\
\begin{itemize}
 
        \item \textbf{Reinforcement learning from human feedback (RLHF)}: Foundation model providers can use RLHF to fine-tune the foundation model's behaviour and produce more accurate and responsible responses. RLHF allows humans to provide feedback on the quality of the responses and uses this feedback to adjust the model's parameters. The foundation model is then trained to maximise the reward it receives from human feedback, which can improve its accuracy and RAI-related qualities over time.
        \item \textbf{Adversarial testing}: Adversarial testing can be done with domain experts to test the accuracy and RAI-related properties through adversarial examples which are designed to deceive or mislead the model into producing incorrect or irresponsible responses. By testing the model in this way, developers can identify vulnerabilities or weaknesses in the model's decision-making process and take steps to improve its accuracy and RAI-related properties. For example, Open AI engaged over 50 experts from various domains to adversarially test their GPT model~\cite{openai2023gpt4}.
        \item \textbf{Data governance measure}: Data governance measures are necessary to examine the reliability of data sources and data biases.

\end{itemize}

\subsubsection{System-level risk control}\hfill\\

\begin{itemize}
\item \textbf{Through APIs}
    \begin{itemize}
        \item \textbf{Foundation model via APIs}: To prevent harmful dual-use and enforce the shared responsibility in the AI value chain, the foundation model provider should impose restrictions on their usage and prevent users from getting around of restrictions through unauthorised reverse engineering or modification of the system design. The foundation model can be provided as AI services on cloud servers so the interactions can be managed through API controls. 
        \item \textbf{Continuous risk assessment}: Continuous risk assessor continuously monitors and assesses the AI systems based on AI risk metrics to prevent the misuse of the AI systems and to ensure the trustworthiness of the systems.
        \item \textbf{Guardrails}: Guardrails can be set up to ensure the outputs are compliant with RAI rules and policies. Prompt refusal involves training the foundation model to identify and refuse inappropriate or harmful tasks, e.g., refuse to generate responses that contain violence promotion or hate speech. Prompts can be appended/generated or modified to make the prompts or the responses more accurate or responsible. 
    \end{itemize}

    \item \textbf{Explainer}
    \begin{itemize}
        \item \textbf{Think aloud}: The think aloud pattern can be used to disclose the decision-making process, such as the intermediate steps like prompt pattern implementation and verification/validation. This design can help build human trust in the system, but it may sacrifice data privacy. 
        \item \textbf{Think silently}: To protect business data and intellectual property, the details of intermediate process might not be revealed.
    \end{itemize}
    \item \textbf{Verifier}: A verifier can verify or modify the responses, and provide feedback.
    \begin{itemize}

            \item \textbf{Human verifier}: A human expert can review and verify the responses returned by a foundation model-based system before they are sent to users. This type of verifier can lead more accurate and responsible responses, but it can also be more time-consuming, expensive, and subjective/biased than automated verifiers.
            \item \textbf{AI verifier}: An AI verifier can use machine learning or other types of AI solutions to identify inaccurate or irresponsible responses ~\cite{fu2023gptscore}. For example, a blockchain smart contract can automatically verify and enforce the rules in an immutable and transparent way. However, it can be costly and limited by blockchain performance. This can have more accurate and responsible responses over time, but it may be limited by the quality of the knowledge data used. Also, an AI verifier may not be able to explain its decisions and may be prone to errors or bias if the knowledge data is incomplete or biased.
            
            \item \textbf{Hybrid verifier}: This type of verifier uses a combination of the above types of verifiers to ensure the trustworthiness of responses. This can be advantageous over any individual verifier, but it can also be more complex and resource-intensive to implement and maintain.
        
    \end{itemize}
    \item \textbf{System audit}
    \begin{itemize}
        \item \textbf{RAI black box}: By recording critical data in an immutable data ledger, the RAI black box allows for accountability analysis after incidents. This includes data such as the input and output of foundation models and small AI models, the versions of foundation models and small (distilled) AI models, etc.
        \item \textbf{Co-versioning registry}: To ensure traceability, the co-versioning registry  can be applied to co-version the AI components, such as foundation models, fine-tuned models, or distilled small models. 
        \item \textbf{Standardised reporting}: The standardised reporting can be used to inform stakeholders (such as regulators and users) about the development process and product design of AI systems, such as AIBOM information about the foundation models and data/model cards.
        \item \textbf{AI bill of materials (AIBOM)}: All software components including the foundation model procured from third parties can be associated with an AI BOM that records their supply chain details, which can include RAI metrics or verifiable RAI credentials. This procurement information can be maintained in an AIBOM registry. 
    \end{itemize}
\end{itemize}

\section{Threats to Validity}
According to the adopted methodology guidelines \citep{keele2007guidelines, threats_to_validity}, threats to validity may be introduced during the literature review process. In this section, we discuss the identified threats to validity and the strategies we employed to minimise their influences.

\textbf{Construct Validity}: The threat to construct validity would be the incompleteness of paper searching. We did not propose the key terms and perform a complete systematic literature review, but selected a seed paper and conducted forward snowballing. This strategy may result in a limited pool of retrieved studies. However, the seed paper is a seminal paper in the field and the most relevant studies can be collected through snowballing. 

\textbf{Internal Validity}: The threat to internal validity may be raised by publication bias and insufficient sample sizes. 
Papers reporting significant or positive findings often have a higher likelihood of acceptance compared to those with null or negative results. We conducted a thorough review of all included studies to mitigate the impact of this potential threat.

\textbf{External Validity}: The threat to external validity may be introduced by the restricted time span in our research protocol. We chose a seminal paper from Stanford University as the seed for paper searching and selection. Since the paper was published in 2021, we excluded the studies before 2021 at the first place, which may affect the generalisability of our result. However, we remark that the selected primary studies are closely related to the theme of foundation model based systems. 

\textbf{Conclusion Validity}: The threat to conclusion validity may be caused by the bias in study selection and data extraction phases. Specifically, the selection and extraction situation may be influenced by researcher's expertise. Four independent researchers were consulted to reduce the biased influences in this procedure.

\section{Conclusion}
In this paper, we propose a taxonomy of foundation models and foundation model based systems. Our taxonomy captures the key characteristics of foundation models, and the major design decisions for the architecture of foundation model based systems. This taxonomy is intended to help with important architectural considerations about the performance and RAI-related properties of foundation model based systems. 
In our future work, we plan to collect a list of design patterns for building foundation model based systems.

\input{main.bbl}



\end{document}